# Exploration, inference and prediction in neuroscience and biomedicine

Danilo Bzdok [1, 2, 3, *] & John P. A. Ioannidis [4, 5, *]

[1] Department of Psychiatry, Psychotherapy and Psychosomatics, RWTH Aachen University, 52072 Aachen, Germany
[2] JARA, Translational Brain Medicine, Aachen, Germany
[3] Parietal Team, INRIA, Neurospin, bat 145, CEA Saclay, 91191 Gif-sur-Yvette, France
[4] Meta-Research Innovation Center at Stanford, Stanford University, Stanford, California, USA
[5] Departments of Medicine, of Health Research and Policy, of Biomedical Data Science, and of Statistics, Stanford University, Stanford, California, USA

*Correspondence: danilo.bzdok@rwth-aachen.de (Twitter: @danilobzdok) and jioannid@stanford.edu

**Abstract**

The last decades saw dramatic progress in brain research. These advances were often buttressed by probing single variables to make circumscribed discoveries, typically through null hypothesis significance testing. New ways for generating massive data fueled tension between the traditional methodology, used to infer statistically relevant effects in carefully-chosen variables, and pattern-learning algorithms, used to identify predictive signatures by searching through abundant information. In this article, we detail the antagonistic philosophies behind two quantitative approaches: certifying robust effects in understandable variables, and evaluating how accurately a built model can forecast future outcomes. We discourage choosing analysis tools via categories like 'statistics' or 'machine learning'. Rather, to establish reproducible knowledge about the brain, we advocate prioritizing tools in view of the core motivation of each quantitative analysis: aiming towards mechanistic insight, or optimizing predictive accuracy.

> '[Deep] neural networks are elaborate regression methods aimed solely at prediction, not estimation or explanation.'
> Efron & Hastie [1, p. 371]

**The emergence of richer datasets alters everyday data-analysis practices**

There is a burgeoning controversy in neuroscience on what data analysis should be about. Similar to many other biomedical disciplines, there are differing perspectives among researchers, clinicians, and regulators about the best approaches to make sense of these unprecedented data resources. Traditional statistical approaches, such as null hypothesis significance testing, were introduced in a time of data scarcity and have been revisited, revised, or even urged to be abandoned. Currently, a growing literature advertises predictive pattern-learning algorithms hailed to provide some traction on the data deluge [2, 3]. Such modeling tools for prediction are increasingly discussed in particular fields of neuroscience [for some excellent sources see: 4, 5-9]. Ensuing friction is aggravated by the incongruent historical trajectories of mainstream statistics and emerging pattern-learning algorithms – the former long centered on significance testing procedures to obtain p-values, the latter with a stronger heritage in computer science [10-12]. We argue here that the endeavor of sorting each analysis tool into categories like 'statistics' or 'machine-learning' is futile [13, 14].

Take for instance ordinary linear regression, as it is routinely applied by many neuroscientists. The same tool and its underlying mathematical prosthetics can be used to achieve three diverging goals [15, pp. 82-83, 16, ch. 4.12]: a) *exploration,* to get a first broad impression of the dependencies between a set of measured variables in the data at hand, b) *inference,* to discern which particular input variables contribute to the target variable beyond chance level, and c) *prediction,* to enable statements about how well target variables can be guessed based on data measured in the future.

Confusion can arise because it is the *motivation* for using linear regression that differs between these scenarios. The mathematical mechanics underlying model parameter fitting are indistinguishable. Taken more broadly, instead of attaching labels of opposing camps to each analysis tool, it would be more productive, we would argue, to focus on the *desired goal* of a specific quantitative analysis. The goal, rather than the choice of a particular tool, is the major factor that ultimately determines what statements can confidently be made about brains, behavior, or genes, or for that matter – any other question of interest.

**Exploration, inference, prediction: A typology of different modeling goals**

The initial description of a correlative relation in brain data is a common first step in many research projects. An important distinction arises when deciding on how to venture into identifying reproducible findings in quantitative analysis. *How* a particular analysis tool is used in a certain application domain may often be more important than *which* class of analysis tool is chosen.

a) Exploration of *correlative* associations: In various studies, a straightforward approach to charting candidate associations in brain data is Pearson's correlation (without computing p-values). A simple statistic is thus computed between two series of measurements for *descriptive* purposes. As one concrete example, this analysis can quantify the relationship between amygdala activity measured in an fMRI experiment and some behavioral response. Such tentative data exploration can also be done in situations involving one input and one output variable by fitting a linear regression to the data. In these informal settings, the modeling goal is limited to a *descriptive, correlational summary* of the raw data that one happened to observe. Estimating linear-regression parameters alone does not license the importance of certain variable relationships (i.e., inference). Neither does a fitted linear regression itself declare whether these variable relationships hold up for other individuals or future data points (i.e., prediction).

b) Inference of *statistically significant* (and possibly causal) associations: Another goal is to try to isolate the specific contributions of single variables, to uncover how the observed response depends on each particular measurement. This is a common agenda in many well-controlled experimental designs. For instance, those looking into the effects of gene-knockout in mice or clinical trials examining the impact of a specific treatment in patients. Historically, this type of deductive reasoning has often drawn on null hypothesis significance testing (NHST). The framework however is sometimes ill-suited and frequently misunderstood [17-19]. As an alternative to NHST, one may draw formal inference by means of false discovery rate (FDR), Bayesian posterior inference, or other tools [1, ch. 3 and 15]. Inferences also need to take into account various biases [20] to avoid making claims that represent false positives (in the NHST framework), underestimated FDR (in the FDR framework), or exaggerated posterior parameter distributions (in the Bayesian framework) [1, ch. 3, 21, ch. 18.7]. Much debate has emerged about what inferential statements about relevant variable contributions mean [10, 13, 22], and how significant associations tends towards the holy grail of uncovering causal influences [23].

c) <u>Generalization of *predictive* associations</u>: One way to substantiate the explored correlations or inferred significance statements is by verifying whether these quantitative relationships still hold up in other data points or new individuals. This goal common to many observational, naturalistic, and prospective epidemiological studies. For instance, increasingly, predictive pattern-learning algorithms are used to derive the behavioral response of individuals from whole-brain neural activity or derive health risk from genomic profiling [cf. 24, 25, 26]. Predictive modeling can also be carried out by standard linear regression. Several fields of clinical medicine have already accumulated a large literature of predictive scores and tools [27, 28]. Currently, usage of predictive approaches lacks standardization and few are rigorously validated [29]. Even fewer are evaluated for replication in different settings and groups of individuals [30]. Increasingly complex predictive models use hundreds and thousands of parameters and/or try to benefit from non-linear interactions in extensive data, like electronic health records [31]. Notably, it has so far rarely been shown that accounting for complex non-linearity in "big" medical data has considerably improved the predictive performance. The low success rate is perhaps partly due to the still insufficient sample sizes or to limited quality of the measurements [7, 20, 32].

To be clear, exploration, inference, and prediction are not strictly mutually exclusive. Rather, quantitative investigations often involve a combination of the three approaches, prioritized to different degrees. In many neuroscience domains that are starting to amass "big data" predictive pattern-learning algorithms are becoming popular alternatives to classic linear-regression applications [2, 3]. Such algorithmic tools include support vector machines, random forests, or artificial neural network. Regardless of whether linear-regression approaches or pattern-learning algorithms are used, the main goal of the prediction enterprise is to put the built model, with already estimated model parameters, to the test in some independent data [21, ch. 7]. In this analysis regime, the investigator wishes to achieve the highest-possible forecasting performance. She is not necessarily worrying about how the model works or whether its fitted parameters carry biological insight.

**Inference and prediction serve distinct goals**

*Scientific insight* has been a primary focus of the statistical methodology traditionally used in fields like psychology, experimental neuroscience, and evidence-based medicine assessments. The underlying inferential approach is particularly suited for asking questions such as, 'Which specific gene location *contributes to* or *has an effect on* a behavioral trait?' Somewhat counterintuitively, in many cases, genetic variants identified via such an inferential approach may not serve best to detect *whether* somebody has that behavioral trait or not [33, 34]. This is because modeling for prediction typically asks a more *heuristic* type of question, 'Which gene locations are *collectively useful* to *distinguish* individuals with or without the behavioral trait?' Finding answers to this latter type of question follows the perhaps more superficial agenda of prioritizing successful recognition of *any data relationships* that are able to derive the specified outcome in independent individuals. Such predictive approaches put less emphasis on *mechanistic insight into the biological underpinnings of the coherent behavioral phenotype* (Table 1).

Inferring new scientific insight is often about answering questions such as 'which input variable within a given dataset is an important contributor to the outcome?' (or, 'a relatively important contributor, compared to other input variables'?) Ideally, this modeling regime *aims at mechanistic understanding of the impact of the input on the target variable*. The investigator is interested in understanding the way in which an outcome y is affected by a change in the input variables $X_1, ..., X_p$. To put it more mathematically, with X denoting the measurement vector $X_1, ..., X_p$, she wants to know 'how y changes as a function of X' [35, p. 19]. Consequently, *inferential data analysis becomes hard if the statistical model is a black box*. Further, inferential statements about individual measurements of brain phenomena have their best chance of being reproducible if derived in the context of careful experimental controls (e.g., randomized trials in clinical assessments). Importantly, however, many, if not most, questions of interest cannot even be addressed using randomization [cf. 1, epilogue].

Quantitative analyses that strive to *mechanistically explain the inner workings of brain phenomena* have a different epistemic value than that of research aimed to *model brain phenomena for the goal of accurate future predictions.* In the prediction case, the investigator wants to extract knowledge about *regularities*, by sieving through constellations of candidate patterns (and possibly very complicated ones) [2, 3]. Prediction accuracy is the core metric to capture how well the overall quantitative model, that is, the collection of fitted parameters, can *emulate* a high-level abstraction of mechanisms in nature. The predictive approach thus asks, 'How well can the built model reproduce the studied brain phenomenon that has been quantitatively captured in the measurements?'

The priority to maximize prediction performance may require exploitation of more complicated non-linear relationships in brain data, in contrast to widely adopted linear modeling. Recognizing complex relationships between variables is something that many black-box pattern-learning algorithms are particularly good at. The more transparent linear-regression approaches have served well in neuroscience and medicine, and are arguably epitomized in the successful era of genome-wide association studies (GWAS) [36]. By contrast, the data-led identification of predictive principles from non-linear relationships between variables has a strong legacy in the machine-learning community [10, 37, ch. 1.2].

A contrast between modeling goals lies in the readiness of non-linear predictive models to capture and capitalize on higher-order interactions among variables. Complex variable-variable(-variable-...) interactions are probably common in brain phenomena. However, to best "see" these higher-order interactions, the data need to be measured with little noise. When adequate data are available, more sophisticated analysis tools are generally advantageous in cases of higher-order variable interactions. Some non-transparent pattern-learning algorithms, capitalizing on non-linear interactions, have frequently ranked among the top solutions in international data-analysis competitions involving a diversity of challenging data types (e.g., http:/www.kaggle.com). In brain research, there is always greater accuracy of measurements and more complete capture of the variables that drive higher-order interactions. Thus, advanced pattern-learning algorithms may eventually outperform linear models even more often than is currently the case. Importantly, though, the superiority of modeling complex patterns over simple linear approaches should not be taken for granted, and merits case-by-case evaluation. Altogether, compared to modeling for inference, the predictive analyst may favor tools extracting regularities from data in a way that is advantageous for prediction accuracy. High forecasting accuracy is favored even if opaque to human intuition, with "deep" neural-network algorithms offering an extreme example of such tools.

Besides challenges in parameter interpretation, predictive tools are typically less suited to detect causal relationships in data [23]. Nevertheless, a useful predictive model with high accuracy may be built based on measurements which are expected to have little causal relation to the outcome of interest. For instance, it has been acknowledged that 'Neuroimaging studies per se [...] only provide insights into neural correlates but not into neural causes of cognition' [38]. Neuroimaging measurements such as fMRI are only indirectly related to the dynamic activity changes in neuronal assemblies underlying cognitive processes. However, such signals carry intermediate information that can serve for accurate predictions of inter-individual differences in cognition such as propensity to attentional lapses, general intelligence, or health status [39, 40].

To recapitulate, we emphasized two types of motivations that could drive a specific scientific inquiry: 'providing insight', for the purpose of inference, or 'accurately modeling the world', for prediction. The inferential regime prioritizes statements about the *relevance of each individual input variables*. The predictive regime instead prioritizes the *relevance of the model's output* for precise forecasting. Predictive modeling describes what '*does*' happen. Prediction often does not equally well address the question of '*how*', and may be less apt for the question of '*why*'. Additionally, prediction is not always feasible and may remain mediocre in certain applications, despite recent technical advances in data analytics. These considerations encourage trade-offs between model transparency for easy interpretability and model complexity that would enable predicting particularly complicated relationships (Figure 1). One could make the case that there are some brain phenomena that are so complicated such that impenetrable predictive pattern-learning algorithms may be all neuroscientists can hope for [cf. 22]. Moreover, accelerating data aggregation and wider availability of computation power are opening a 'shortcut' path to useful outcome predictions, circumventing the traditional milestone of mechanistic discovery as an essential step towards effective predictive capabilities.

**Implications for clinical brain research**

Many clinical studies in brain research set out to identify variables that are statistically significantly associated with a disease. This includes significant differences in specific brain regions, their neural activity or anatomical abnormality, connections between brain regions, gene variants, and more. Deviations in such measurements in patients, however, may not always be best-possible choices for building successful *predictive* approaches [20, 41, p. 185]. This is perhaps not too surprising, given that certain questions beg modeling for the *inference* goal. For instance, 'Which particular demographic indicator, ethnic background, or clinical parameter is robustly associated with adverse reaction of patients to a drug?'. The context of *predictive* modeling begs a different question at the heart of the study, even when using the same statistical technique. For instance, 'How well can we know in advance the risk of a particular patient for an adverse reaction to that drug?'. Predictive modeling regimes, we would argue, provide a natural path towards clinical relevance, by immediately acting on clinical endpoints [42]. In fact, an official report of the American Statistical Association (ASA) emphasized that 'Statistical significance is not equivalent to scientific, human, or economic significance. Smaller p-values do not necessarily imply the presence of larger or more important effects, and larger p-values do not imply a lack of importance or even lack of effect.' [17].

Modeling for inference and prediction are two different tasks. Increasing this awareness will probably foster new research directions. Centering on clinical endpoint predictions can complement the quest for identifying the biological causes of disease. Historically, in research on the neural and genetic basis of brain disease, a prevailing philosophy has been to progress in two consecutive steps: discovery of new pathophysiological mechanisms, which are then used as a stepping stone to designing new targeted treatments [43]. Yet, one might argue, after >50 years of biological research on the brain aimed at inference, there are relatively few definitively established etiopathological pathways. Neither are there many reliable biomarkers for most mental disorders [44].

Even in the ideal case of brain diseases caused by a single gene with considerable penetrance, such as in 22q11.2 deletion linked to schizophrenia risk [45] and expansion of CAG triplet repeats linked to Huntington's disease [46], certain clinical endpoints can profit from patient-tailored predictive approaches. All individuals with such a genetic variant carry an escalated risk of developing the disease. However, various inter-individual differences can still arise, including the timing of symptom onset, the constellation of symptoms displayed, disease severity, clinical trajectory, and treatment response. These clinical scenarios illustrate the distinction between the pursuit of scientific insight and the wish to forecast patient-specific disease manifestations – aiming at elucidating disease-causing biological mechanisms or prognostic value with relevance for medical care. Without doubt, there potentially are immediate gains of the pragmatic intention to search signatures in complex data that can be exploited to predict clinical endpoints. Such research program does not conflict with or question the value of the longer-term endeavor to understand the primary biology of brain diseases.

Predictive approaches are increasingly adopted, recommended, and even expected by policymakers [47, 48]. However, there are several requirements before they can be considered suitable for wide application in real-world clinical settings (Box 1). Beneficial conditions for successfully translating new predictive approaches into clinical practice include the following**:**

i) The input variables for the predictive approach should be unambiguously defined as well as measured in a straightforward and standardized way.

ii) The prediction performance needs to be better than what can be achieved by already existing clinical means for diagnosis and monitoring.

iii) Accurate predictions need to be carefully validated in diverse settings [49]. It is important to accommodate variability that results from contextual factors such as circadian rhythm, menstrual cycle, and periods of stress.

iv) The predictive approach must also show reproducibility in different groups of individuals and different ethnicities that did not contribute to model building. In analogy to drug treatments, a candidate predictive model can be found, for instance, to work better in males than females or be less effective in the elderly. Drug treatments can also have adversary outcomes in individuals with specific genetic profiles [cf. 7].

v) Predictive successes can only result in better patient management and clinical outcomes if effective interventions are available. In Alzheimer's disease, for instance, a major current effort is directed to improving disease prediction years prior to symptom onset. Translating such prediction to better clinical outcome, however, would depend on whether treatment interventions that can effectively leverage such earlier diagnosis do exist.

vi) Successful predictive models that are easy-to-use and transparent are likely to be adopted more readily by the medical community. Health professionals will probably avoid complex modeling approaches that are harder to interpret, require extra training or depend on hard-to-get information.

vii) Randomized clinical trials may need to certify the utility of a new predictive approach for patients [50, 51]. This important cornerstone of evidence-based medicine will most likely continue to bolster clinical guidelines in the "big data" era.

Finally, we outline various obstacles in the journey towards establishing predictive approaches for clinical management and intervention:

i) When using medical data, strong non-linear effects have seldom been explicitly modeled or reported [52]. Even if complex interactions exist between measured variables, they may be difficult to extract from today's datasets, particularly those of still limited sample sizes [20]. Consequently, simple, less data-hungry predictive approaches are likely to remain among the go-to choices in many clinical settings. Elaborate predictive pattern-learning algorithms often cannot yet be used to their full potential, let alone "deep" neural-network algorithms [cf. 53].

ii) It is often hard to know the optimal sample size for a particular prediction-oriented clinical research program beforehand. This limitation stands in contrast to the availability of power calculations in classical statistics. Reasons include the unknown complexity of the aspired prediction function, amount of relevant input variables and noise in the data [20, 54, 55, p. 124].

iii) A small signal-to-noise ratio plagues various forms of medical data. Examples of noisy measurements include read-outs of histone modifications in genomics and brain activity changes scanned using functional MRI, EEG, or MEG. As a rule of thumb, the more complex the predictive model, the higher its susceptibility to random variation in the data. Hence, it is trickier for advanced pattern-learning algorithms to identify reproducible relations among the measured variables.

iv) Similarly, flexible predictive pattern-learning algorithms with high capacity are more prone to overfitting idiosyncrasies in the data [56]. Thus, the various "bells and whistles" of many of the sophisticated predictive approaches need to be chosen in a principled fashion [52]. These considerations underscore the need for reproducible modeling practices as a core activity in brain research [cf. 57].

v) The lacking transparency of predictive approaches going beyond mainstream linear modeling is a particular concern that can erode the trust needed for implementation in clinical practice [47, 52]. Indeed, skewed or wrong predictive approaches can systematically inflict harm by driving poor decision making [58].

vi) Because of methodological constraints, much clinical brain research may not directly target real-world settings. Rather, clinical studies routinely enroll patients based on stringent exclusion criteria, such as medication use or common comorbidities. These study designs may impede our ability to make predictions in realistic clinical settings. For instance, assessing the effectiveness of drugs or other treatments is particularly hindered when it comes to groups of patients that are relatively rarely recruited in clinical studies, such as children and the elderly [59].

vii) Electronic health records are soon likely to provide rich resources to build effective predictive approaches. Yet, standardized health records involving large samples of patients are still scarce. Additionally, a bias towards sicker people has been noted in the few existing studies using such patient data provided by medical institutions [30, 60].

**Concluding remarks and future perspectives**

The advent of "big data" in neuroscience and biomedicine started transforming many important sectors. In the 21st century, large-scale data aggregation, catalyzed by new modes of data dissemination and open science [61], has reached an unprecedented scale. Yet, it remains unclear whether these emerging opportunities also prompt a deeper revision of the traditional "value system" pertaining to scientific evidence. The data-rich neuroscientist can ask many new questions that could probably never be addressed quantitatively before. We encourage investigators and clinicians to re-think data analysis in the context of a repertoire of modeling goals (see Outstanding Questions). *Choosing a data-analytic strategy for a research question at hand should not be a matter of tradition, habit, or taste*.

It is worth reiterating that a specific analysis tool can serve multiple modeling goals. Linear regression, for instance, has been often used for exploratory summaries of possible relationships among measured variables. The same tool, however, can be used for inferring the most relevant mechanistic candidates among the measured variables, as well as predicting outcomes by applying the built model to new data points. Conversely, many machine-learning algorithms have a long-standing track record in serving the predictive goal. Yet, despite the increased complexity of many of these algorithmic tools, they can also be partly used towards the aim of data exploration, or even inference to isolate individually important input variables.

More broadly, like any scientific method, modeling for inference or prediction both come with certain strengths and weaknesses [19, 34, 62, 63]. Inferential modeling has been an established practice for decades [50, 64]. In contrast, the most effective use cases still need to be identified for deploying predictive approaches in neuroscience and personalized medicine. Ultimately, deducing scientific insights and making pragmatic predictions are intimately related, but also differ in important ways.

**Acknowledgments**

We are grateful to Jérémy Besnard-Lefort, Avram Holmes, Hannah Kiesow, Timm Poeppl, Marc-Andre Schulz, Bertrand Thirion, and Thomas Wiecki for insightful comments on a previous prevision of the manuscript. We thank three anonymous reviewers for the many insightful comments.

Dr. Bzdok is funded by the Deutsche Forschungsgemeinschaft (DFG, BZ2/2-1, BZ2/3-1, and BZ2/4-1; International Research Training Group IRTG2150), Amazon AWS Research grants (2016 and 2017), as well as the START-Program of the Faculty of Medicine (126/16) and Exploratory Research Space (OPSF449), RWTH Aachen.

**Table 1: The inference-prediction continuum of modeling goals (cf. Figure 1)**

| Inference <----------------------------------------------------> Prediction | |
|---|---|
| Commonly used tools for inference goals: Null hypothesis significance testing to compute p-values for specific variables. Tools to this end include, for example, ANOVA, t-test, or $\chi^2$-test. Increasingly popular alternatives include false discovery rate and Bayesian posterior inference, as well as certain pattern-learning algorithms (e.g., feature importance scores from random-forest algorithms) | Commonly used tools for prediction goals: Validation schemes to compute prediction accuracy of the built model as a whole. Exemplary tools include support vector machines, random-forest algorithms and other ensemble and boosting techniques, the rapidly evolving "deep" learning algorithms, as well as ordinary and penalized linear regression |
| Theory-guided: Candidate variables are often hand-picked by the investigator, in a targeted fashion based on existing substantive knowledge. Research questions are explicitly articulated before data collection in a carefully controlled experiment. The chosen variables are evaluated by an often simple, inflexible model that ideally, is pre-specified by the investigator before seeing the data; but data dredging, and thus high false-positive rate is common in practice | Pattern-guided: A large and diverse array of "found" variables is typically considered in the statistical analysis in a heuristic data-led fashion. It can be unknown how the data were generated, and the exact research question may be detailed as the data are being analyzed. The adaptive, sometimes very flexible model extracts a general prediction rule directly from the data themselves |
| Explainable narrative: Statements about the specific contribution of individual input variables are the priority. Such claims of variable relevance are often more readily available in simple linear regression models. Accordingly, these models tend to be preferred in this context, such that every single parameter can be cleanly attributed its share of the explained variance. Usually, the meaning of each parameter should be readily understood and hence the model often allows for a simplified narrative; statements are centered on single parameters, rather than the prediction performance of the collective model parameters | Opaque black box: While simple linear-regression models may perform reasonably well in terms of predictive power, if the goal is to maximize prediction accuracy, it is often beneficial to exploit complicated non-additive associations in the data. In many real-world situations, the target variable depends on the input variables in convoluted ways, which can hinder assigning to single input variables a clear relative contribution to the output; model parameters are often treated as instrumental intermediates to achieve high prediction performance, without necessarily aiming to assign specific meaning to each parameter estimate *per se* |
| Formally justified: Many traditional analysis techniques were rigorously characterized by mathematical theory; simple linear models lend themselves well for theoretical model criticism, and carry well-understood modeling limits; another benefit is the typically lower computational load | Empirically justified: Predictive models can be explicitly and quantitatively evaluated by applying the entire set of model parameters to unseen independent, newly generated, or future observations or individuals; formal performance guarantees are often challenging; these models are closely related to more computationally demanding cross-validation, bootstrapping, and other resampling schemes |
| Data-efficient: Many classical statistics methods were designed long ago to handle data that are scarce, as well as laborious and expensive to collect | Data-hungry: Compared to classical statistics methods, many complicated predictive approaches require more data, especially when complex non-linear relationships are modeled, and more hyper-parameters need to be tuned; comparably more data are also needed if each observation tends to have many input variables, and random noise is expected to be prominent (e.g., medical data) |
| Problem-tailored: Each approach is designed to solve a particular data-analysis question, typically based on a | Versatile: Approaches are devised to provide useful solutions to various types of data and data-analysis |

| problem-specific probability model of how the data are believed to have come about | questions |

**Box 1: Stages of translating predictive approaches in brain research into practice**

1. Model building: To fit the parameters of the chosen predictive model, one first needs empirical measurements from the brain systems of interest. One common preparatory analysis is probing variable-variable relationships using pairwise correlation plots. Another is estimating genetic relatedness between the participants using principal component analysis of the genomic profiles. In behavioral experiments in animals or humans, exploratory data summaries can identify collinearity in response times. Such collinearity in response times foreshadows hindered statements about the relevance of individual experimental conditions (i.e., inference); but hardly affects forecasting condition response latencies in new participants (i.e., prediction).

2. Internal validation: These procedures guard against overly optimistic modeling performances. Internal validation procedures, unlike external ones (point 3), do not require new, independent data and are based only on the original subject sample or dataset that was used during model building [65]. Cross-validation and bootstrapping are resampling methods [21, ch. 7] that can estimate metrics of model quality [47], such as expected prediction accuracy on future data, uncertainty of parameter estimates, and variability of prediction errors. Indeed, 'working scientists often find the most interesting aspect of the analysis in the lack of fit rather than the fit itself' [16, p. 92]. Yet, inter-individual variability may still be under-appreciated by using these internal validations alone [24].

3. External validation: For further validation, predictive associations identified from the original subject sample or dataset need to be checked in other individuals or in datasets measured later [60, 64, 66]. Successful application of a predictive model of disease risk, for instance, requires validation in different groups of individuals [24, 29]. This step is important to combat reproducibility issues [67]. Currently, such model validations are not done as often as they should be [68]. It is however important to comprehensively benchmark the value of each predictive approach for clinicians, policy makers, and clinical guidelines [69]. For instance, external validation may be performed in different geographical areas, time periods, and settings (e.g., secondary vs. primary care). Generally, some authors proposed that 'the most stringent external validation involves testing a final model developed in one country or setting on subjects in another country or setting at another time. This validation would test whether the data collection instrument was translated into another language properly, whether cultural differences make earlier findings nonapplicable, and whether secular trends have changed associations or base rates' [16, ch. 5.3.1].

4. Generalizability and transposability: When testing the predictions of a model on new individuals, the more different these individuals are from the original subject sample, the stronger the test for generalizability [59, 65]. Prediction accuracies are typically lower than in preceding steps. For instance, our ability to predict the clinical utility of drugs tends to be particularly hindered for certain groups of patients, including women, children, the elderly. Common comorbidities are also frequently underrepresented or excluded in clinical studies. Meta-analysis methods can be useful for summarizing and examining a model's predictive performance across different scenarios. Large datasets from multiple studies and electronic health records or registry databases provide promising opportunities for examining the generalizability of predictive approaches [70].

To enhance reproducibility, accurate and complete reporting in studies applying predictive models is imperative. Such reporting is crucial for being able to critically appraise predictive models; to perform acid-test validations of them; to evaluate their impact; and ultimately, to translate them into clinical practice [27, 71].

**Figure Legends**

**Figure 1: The trade-off between model transparency, which allows for scientific understanding, and theoretical model capacity, which affords sophisticated predictions**

Neuroscience and biomedicine had a long-dominating focus on scientific insight using simple and thus transparent models. Such approaches are well suited to work towards the goal of inference on mechanistic understanding. This goal is epistemologically distinct from and sometimes practically incompatible with maximizing predictive power. The pragmatic goal of optimizing predictive accuracy can exploit large datasets even at the cost of opting for black-box models that cannot easily be interrogated. In practice, the actual ratio between transparency and predictability depends on the specific analysis tool being used and the particular dataset at hand. Abbreviations: GLM: generalized linear models; LASSO (least absolute shrinkage and selection operator): a recently introduced constrained regression for high-dimensional data analysis, which is a special instance of GLM.